\documentstyle[sprocl]{article}

\def\Journal#1#2#3#4{{#1} {\bf #2}, #3 (#4)}

\def\be{\begin{equation}}
\def\ee{\end{equation}}
\def\bea{\begin{eqnarray}}
\def\eea{\end{eqnarray}}

\begin{document}

\title{Sub-mm clues to elliptical galaxy formation}

\author{James S. Dunlop}

\address{Institute for Astronomy, Royal
Observatory, Edinburgh EH9 3HJ, UK}


\maketitle\abstracts{ 
There is growing evidence that, at the $S_{850\mu m} < 1$mJy level, 
the sub-mm galaxy population (and hence a potentially 
significant fraction of the 
sub-mm background) is associated with the star-forming Lyman-break 
population already detected at optical wavelengths. However, the implied
star-formation rates in such objects (typically $3 - 30 {\rm M_{\odot}
yr^{-1}}$) fall one or two orders of magnitude short of the level of 
star-forming activity required to 
produce the most massive elliptical galaxies on a timescale $\sim 1$Gyr.
If a significant fraction of massive ellipticals did form the bulk of their
stars in short-lived massive starbursts at high redshift, then they should 
presumably be found among the brighter, $S_{850\mu m} \simeq 10$mJy sub-mm sources 
which are undoubtedly {\it not} part of the Lyman-break population. A first 
powerful clue that this is indeed the case comes from our major SCUBA survey of
radio galaxies, which indicates that massive dust-enshrouded
star-formation in at least this subset of massive ellipticals is 
largely confined to $z > 2.5$, with a mean redshift $z \simeq 3.5$.
While radio selection always raises concerns about bias,
I argue that our current knowledge of the brightest
($S_{850\mu m} \simeq 10$mJy) sub-mm sources detected in unbiased SCUBA imaging 
surveys indicates that they are also largely confined to this same high-$z$ 
regime. Consequently, while the most recent 
number counts imply such extreme sources 
can contribute only 5-10\% of the
sub-mm background, their comoving number density (in the redshift band $3
< z < 5$) is $\simeq 
1 - 2 \times 10^{-5} {\rm Mpc^{-3}}$, sufficient to account for the formation of {\it
all} ellipticals of comparable mass to radio galaxies ($\ge
4L^{\star}$) in the present-day universe.}

\section{Introduction}

The formation mechanism of elliptical galaxies remains a fundamental and
controversial issue in cosmology.  
In current models of galaxy formation
dominated by cold dark matter (CDM), elliptical galaxies arise from the
merging at low redshift of intermediate-mass discs \cite{ba}, and 
some recent data have been interpreted as supportive of the 
implied gradual formation of massive ellipticals at 
relatively low redshift \cite{kaa}.
However, the validity of these analyses has recently been questioned
\cite{to,ji}, 
and other 
observational evidence at low/moderate redshift \cite{bo,no,be}, 
continues to favour a picture in which at least some
massive ellipticals
formed the bulk of their stars in a relatively short-lived
($\le 1$ Gyr) massive starburst at high redshift ($z > 3$) \cite{ji}.
While this 
high-redshift star-formation scenario might only apply to a  
subset of ellipticals \cite{me},
it can be argued that it applies to massive ellipticals 
in general \cite{re}. Moreover, the 
clarification of the link between
black-hole and spheroid mass \cite{ma}
suggests that the hosts of AGN may be more representative of
spheroids in general than previously supposed.

\begin{figure}
\vspace{5cm}
\centering
\setlength{\unitlength}{1mm}
\includegraphics{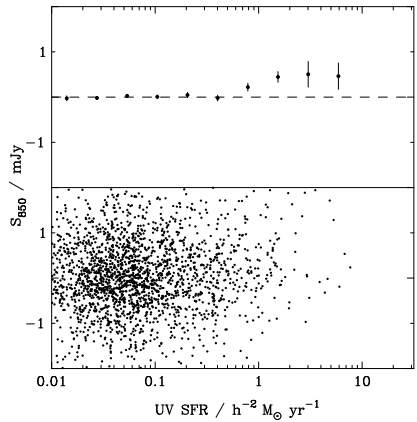}
\caption{A plot of observed 850-$\mu m$ flux density against UV SFR for
known optically-selected starburst galaxies in the HDF. The bottom panel
shows the raw values, while the top panel shows the mean values with
standard errors. A statistically significant sub-mm detection is obtained
for star-forming galaxies with $SFR > 1 h^{-2} {\rm M_{\odot} yr^{-1}}$}
\end{figure}

\section{Observing spheroid formation - optical versus submm}

It has been argued that the formation of present-day galactic
bulges/spheroids has already been observed at optical wavelengths,
through the discovery of the Lyman-break population at $z \simeq 2 - 4$ 
\cite{st}. However, even if 
substantial corrections are made to correct for the effects of
dust, the inferred star-formation rates in these objects are relatively
modest (typically $3 - 30 {\rm M_{\odot} yr^{-1}}$) \cite{pet}. 
Even luminous
Lyman-break objects thus appear to fall over an order of magnitude short of
the high star-formation rates ($\simeq 1000 {\rm M_{\odot} yr^{-1}}$) required 
to construct the stellar populations of the most massive ellipticals 
on a timescale $\le 1$ Gyr.

Direct confirmation of this comes not only from the difficulty experienced 
in detecting individual, unlensed Lyman-break objects with SCUBA \cite{ch}, but
also from the achievement of a statistical detection of the 
bright end of the Lyman-break population in the deep SCUBA image of the HDF
\cite{pea}. The basic evidence for this result is shown in 
Fig.1; Lyman-break galaxies with raw (uncorrected, UV-derived) 
star-formation rates $\simeq 1 h^{-2} {\rm M_{\odot} yr^{-1}}$ 
are detected, statistically, in the 
SCUBA image at flux densities of $S_{850\mu m} \simeq 0.2$ mJy. This number
implies that the ratio of hidden-to-visible star-formation in these objects 
is $5 \pm 1.5$ (consistent both with dust-screen models
\cite{pet}
and with models which assume the first `15 Myrs-worth' of star-formation in 
giant molecular clouds is always essentially invisible at UV wavelengths
\cite{ji}). It also implies that the Lyman-break population is 
probably responsible for a significant, albeit highly-uncertain fraction of 
the sub-mm background ($\sim 25$\%).

This independent check on the hidden:visible star-formation ratio in 
Lyman-break galaxies confirms that optical surveys very rarely
uncover galaxies forming stars at the level of 
$100 -  1000 {\rm M_{\odot} yr^{-1}}$. One possible 
explanation for this is that the formation
of the stellar populations of massive ellipticals is too widely 
distributed in space and/or time to be identified with a very violent event.
Observationally, however, it remains possible that 
such massive starbursts are prevalent at high redshift, but are either 
too dust-enshrouded, or at too extreme $z$ to be detected 
in existing optical-UV drop-out surveys.

In fact, since the first sub-mm detections of the $z \simeq 4$ radio galaxies
4C41.17 and 8C1435+635 five years ago \cite{dua,iva},
it has been clear that such massive dust-enshrouded 
high-$z$ starbursts do at least exist.
These galaxies display 
exactly the properties expected of a young 
massive elliptical , 
with inferred dust-enshrouded star-formation rates 
$\simeq 1000 {\rm M_{\odot} yr^{-1}}$. 
However, the relevance of such extreme sources to the 
general elliptical population has been unclear.

\section{Sub-mm studies of radio galaxies out to $z \simeq 4$}

\subsection{SCUBA photometry of radio galaxies}

These pioneering sub-mm observations have therefore raised the important 
issue of whether all luminous sub-mm sources are largely confined to 
extreme redshift ($z \simeq 3 - 5$). A powerful clue that this 
may be true comes from our major SCUBA study \cite{ar} of radio galaxies spanning 
the redshift range $0 < z < 5$.
Such a survey is potentially biased, being based on an
(arguably) special subset of massive ellipticals. However, it 
does offer the advantages of ready-made 
redshift information and accurately known positions (allowing SCUBA 
to be used in its most sensitive photometry mode).
These advantages have enabled us to deduce that the typical 
level of dust-enshrouded star-formation in radio galaxies grows rapidly
beyond $z \simeq 2$, continuing to rise out to $z \simeq 4$ (Fig.2).

\begin{figure}
\vspace{4.7cm}
\centering
\setlength{\unitlength}{1mm}
\includegraphics{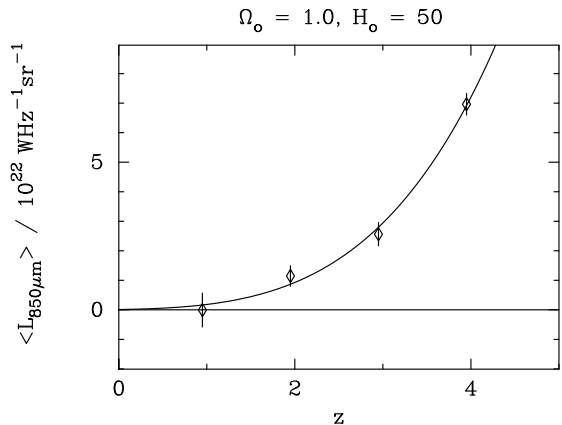}
\caption{The rapid growth of average sub-millimetre luminosity with
increasing redshift found for powerful radio galaxies spanning the
redshift range $0.5 < z < 4.5$.
The datapoints are the result of SCUBA
observations of a sample of $\simeq 50$ radio galaxies, placed in
redshift bins of unit width. The curve has the form $L \propto (1+z)^4$,
and serves to illustrate the dramatic nature of the increase in
characteristic sub-mm luminosity, particularly beyond $z \simeq 2.5$.
This result suggests that the epoch of maximum star-formation in the
massive elliptical hosts of radio sources lies at $z \simeq 3-4$. 
}
\end{figure}

How biased might this result be? Interestingly the median redshift of the
submm detections in the radio-galaxy sample is $z \simeq 3$, 
consistent with that derived for the (cluster-lensed) 
field population \cite{ar}. However, as shown in Fig.2, the average 
sub-mm luminosity continues to rise beyond $z \simeq 3$.
This is because 4 of the 5 most luminous sub-mm sources ($S_{850\mu m} > 5$ mJy) 
in the sample lie at $z > 3$. Could this 
be true for luminous sub-mm sources in general?
 
\subsection{SCUBA imaging of radio galaxies}

Already it is clear that large sub-mm luminosities 
at $z \simeq 3 - 4$ are not simply confined to the host galaxies of
AGN. New, deep, SCUBA {\it imaging} of the regions 
around several of the above-mentioned radio galaxies has revealed even
more submm-luminous  companion sources, at the same 
redshift \cite{ivb}. 
The implication is that high-$z$ radio
galaxies act as signposts towards young clusters in which 
much of the eventual stellar content of the cluster ellipticals is forming
in massive dust-enshrouded starbursts ($\simeq 1000 {\rm M_{\odot} yr^{-1}}$). 
This is consistent with the apparent age and
coeavality of present-day cluster ellipticals. 
Moreover, the faintness of the possible optical IDs of these
very luminous sub-mm sources indicates that this process
is basically invisible at optical wavelengths. 

\section{Bright sources from unbiassed sub-mm surveys}

With the advent of sensitive sub-mm
imaging with SCUBA, unbiassed sub-mm surveys with the potential to
properly quantify the prevalence of massive dust-enshrouded starbursts
at high $z$ can now be carried out. Several such surveys are 
underway \cite{eab,sm}, but it is clear
that reliable source detection, optical/IR identification, and
redshift determination is still in its infancy.

\begin{figure}
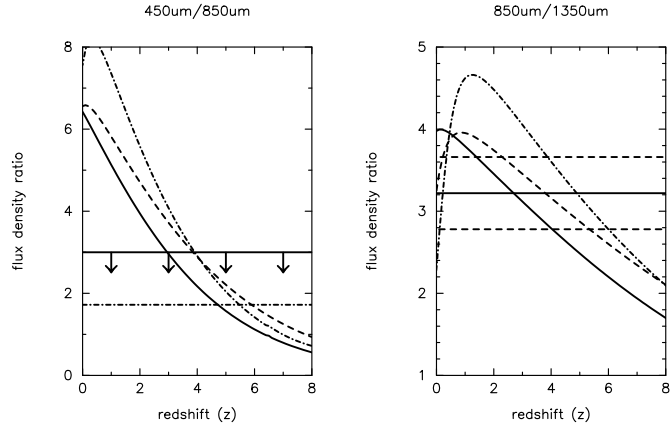

\vspace{5cm}
\leavevmode
\includegraphics{umassfig3a.eps}
\includegraphics{umassfig3b.eps}
\caption{Observed 450/850$\mu$m and 850/1350$\mu$m flux-density ratios
for HDF850.1 compared with what might be expected  as a function of
redshift. The curves show the extreme range of colours for SEDs that
represent dust enshrouded starburst galaxies and AGN - Arp220 (solid),
M82 (dashed) and Mkn231 (dot-dashed). The solid horizontal line in the
left-hand panel shows the existing  upper limit to the 450/850$\mu$m 
flux ratio for HDF850.1. The horizontal lines in the RH panel show
the measured 850/1350$\mu$m flux ratio for HDF850.1 (solid line), and the $\pm
1\sigma$ errors on this ratio (dot-dashed lines).}
\end{figure}

One of the main reasons for this somewhat slow progress is the fact that 
it is only for bright ($S_{850\mu m} > 8$ mJy) SCUBA sources that unconfused
positions can be reliably obtained with, for example, follow-up mm 
interferometry. This is well demonstrated by the effort required to determine
an accurate position for the brightest  source (HDF850.1; $S_{850\mu m} = 7$ mJy)
from our SCUBA survey of the HDF \cite{do}.
Indeed, even with sub-arcsec positional accuracy, it can prove hard to distinguish
between alternative candidate IDs and, as the test-case of HDF850.1
demonstrates, it is often vital to supplement good astrometric
information with SED-based redshift constraints derived from deep
radio$\rightarrow$far-infrared photometry (Fig.3). In this case 
it is hard to escape the conclusion that, as for the radio
galaxies and their companions, this luminous sub-mm
source lies at $z > 3$, in which case its optical counterpart may 
well be too faint/red to be
detected in published optical/IR images of the HDF. 

\subsection{The 8-mJy SCUBA survey}

In an effort to assemble a substantial and unbiased sample of sub-mm
sources of a luminosity comparable to or greater than HDF850.1, 
we are currently undertaking an `8-mJy' 
SCUBA survey covering 400 sq.arcmin. In 
comparison to existing SCUBA surveys this new survey has 
four key advantages. First, its flux limit is sufficiently bright to 
allow follow-up with existing instrumentation (e.g. the IRAM PdB 
interferometer) which should ultimately 
yield an accurate ($\sim 1$ arcsec) position
for every source. Second, there is no a priori reason to expect the
sources to be lensed. Third, the flux limit is sufficiently bright that 
sub-mm confusion is not a problem. Fourth, and scientifically most important,
any sources discovered in this survey must be as bright or brighter in the
sub-mm than the extreme radio galaxies mentioned above. This survey is 
therefore optimised for the detection of starbursts with SFR $\simeq 
1000 {\rm M_{\odot} yr^{-1}}$.
Initial results from this survey indicate a cumulative source count 
of 250$\pm$70 degree$^{-2}$ at flux densities $S_{850\mu m} \ge 10$mJy, and
provide preliminary evidence that these 
luminous sub-mm sources are strongly clustered (Fig.4).

\begin{figure}
\vspace{6cm}
\centering
\setlength{\unitlength}{1mm}
\includegraphics{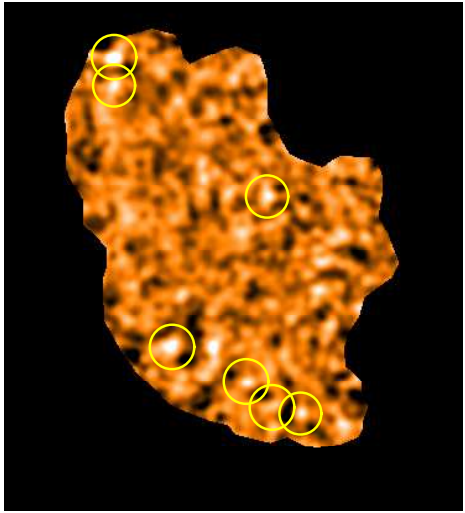}
\caption{Current SCUBA map of the ELAISN2 region covering $\simeq 100$
sq. arcmin. The seven most secure ($\ge 4 \sigma$) sources with $S_{850\mu m} >
9$mJy, are marked by circles. The beam signature, with 
negative sidelobes 45$^{\circ}$
east-of-north, is clearly visible for the brighter sources.}
\end{figure}

\begin{figure}
\vspace{2.8cm}
\centering
\setlength{\unitlength}{1mm}
\includegraphics{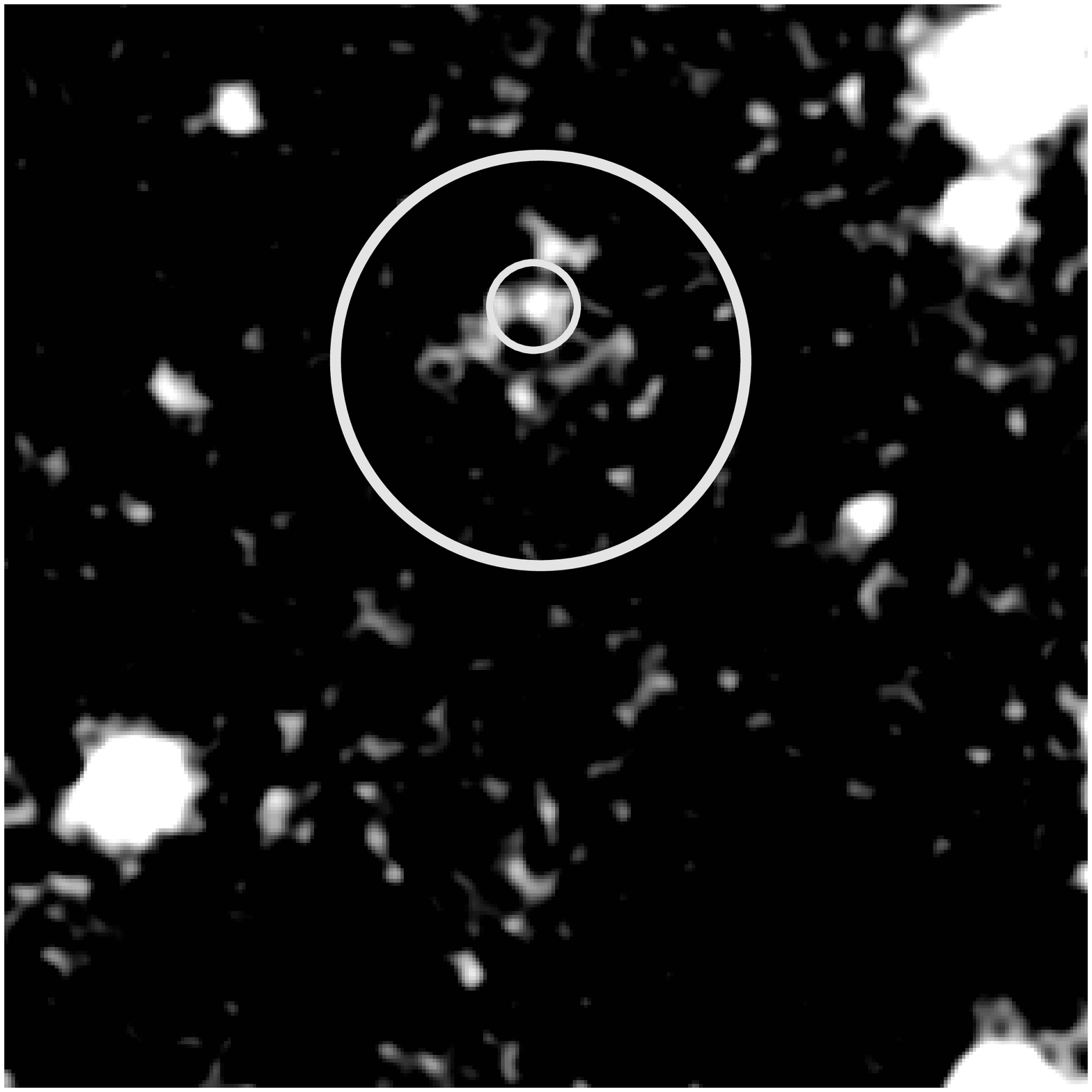}
\includegraphics{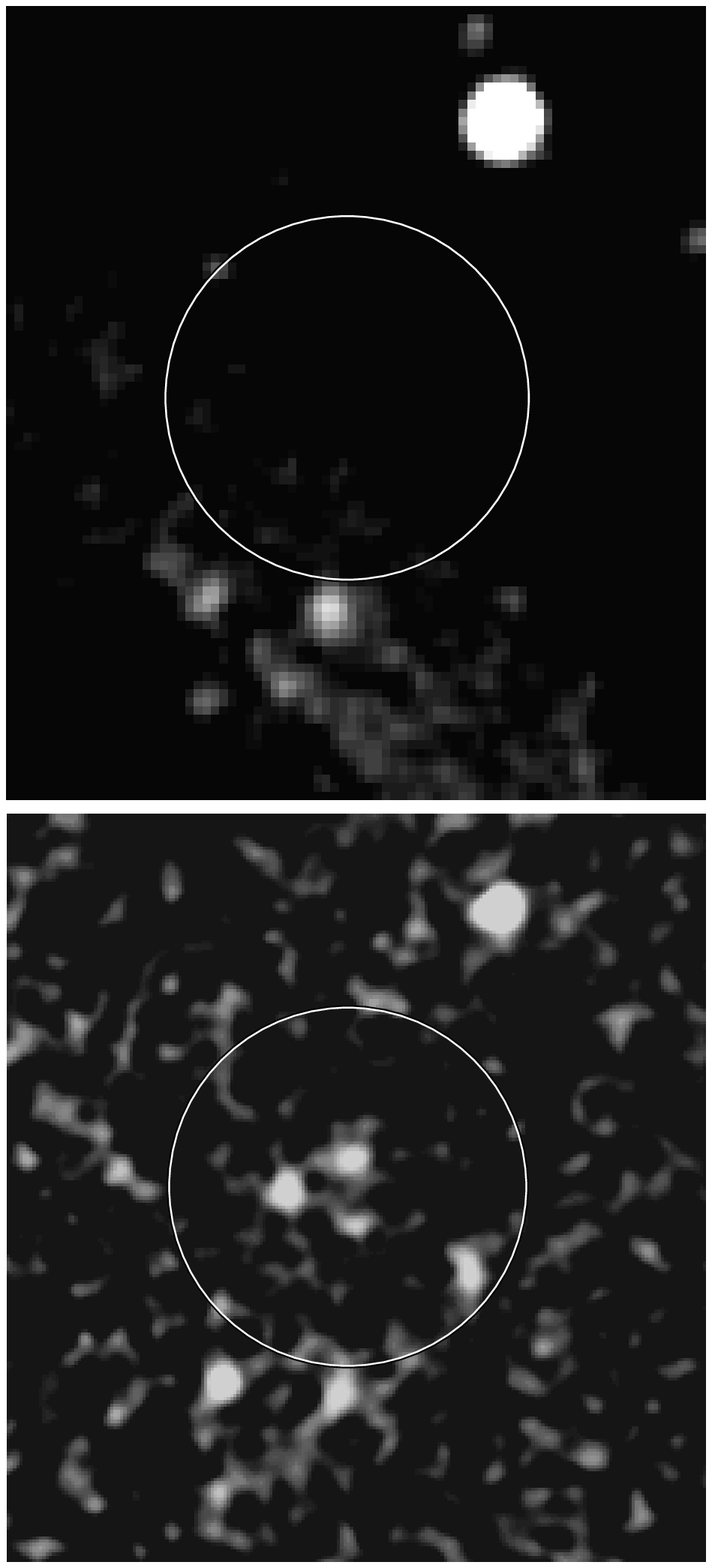}
\caption{Current information on the two most luminous sub-mm sources
detected in the SCUBA 8-mJy survey. The left-hand panel shows a 30 x 30
arcsec region of a deep (6-hr on UKIRT) K-band image of the
field in the vicinity of the $S_{850\mu m} = 11$ mJy source Lockman850.1.
The position of the SCUBA source is marked by the large (10-arcsec
diameter) circle, while the position of the 1.3mm source detected in
follow-up observations with the IRAM PdB interferometer is marked by the
small (2-arcsec diameter) circle.
With this positional accuracy the SCUBA source
can be confidently identified with the brightest of a group of 
compact peaks with $K \simeq 21$, and $R-K > 5$. 
The two right-hand panels show the
position of the even brighter SCUBA source ELAIS850.1 ($S_{850\mu m} =
15$ mJy) superimposed on a 1-hour K-band UKIRT image (centre panel) and
an $R$-band image reaching $R \simeq 26$. We still
await IRAM PdB observations of this object, but 
again the bright SCUBA source is
associated with a group of faint, red knots ($K \simeq 20.5$; $R-K > 5$).}
\end{figure}

The crucial next stage is to determine the nature/redshift of the
bright SCUBA sources in this survey. As a first step
we have been investigating the two brightest sources
detected to date in our 8-mJy survey. These two sources - ELAIS850.1
and Lockman850.1 - have flux densities at 850$\mu m$ of 15 and 11 mJy 
respectively, making them among the very brightest unlensed SCUBA sources
discovered from blank-field surveys so far. 
For Lockman850.1 we have obtained a clear
detection at 1.3mm with the IRAM PdB interferometer, yielding its position
to sub-arcsec accuracy \cite{lu},
and similar observations are also planned for ELAIS850.1. We have also
obtained deep $K$-band images of these sources with UFTI on UKIRT. 
The results of this sub-mm/mm/infrared/optical 
comparison are shown in Fig.5. In both
cases the SCUBA source is associated with a clump of very 
red objects, with $K \simeq 21$ but $R > 26$, and in the case of 
Lockman850.1 the IRAM position ties the SCUBA source to the brightest of these 
clumps. What is so striking about these images is their apparent 
similarity, and in particular the complexity of the sources at $K$. We
are certainly {\it not} seeing either an obscured AGN nucleus, or a
relaxed elliptical galaxy at intermediate redshift. In fact, these images 
are very reminiscent of the complex $K$-band morphologies found for 
radio galaxies at $z > 3$ \cite{van}, reinforcing the connection 
between radio galaxies and the bright sub-mm population in general.

Finally, I note that the one comparably-bright sub-mm source 
reported from the SCUBA survey of the CFRS \cite{eab} has also
been identified with an ERO via IRAM interferometry \cite{ge}, and   
again has SED constraints indicating $z = 2 \rightarrow 5$.
I conclude, therefore, that if attention is confined to the most luminous
sub-mm sources, we find rather little `diversity' in this population. Whether
radio galaxies, cluster companions, or blank field sources, all 
appear to lie at $z > 2.5$, and are associated with (often complex) EROs.

\section{Implications}

Given their surface density, extreme (
$S_{850\mu m} > 10$mJy) sub-mm sources can only 
contribute 5-10 \% of the sub-mm background.
However if this population is indeed confined to a 
relatively narrow redshift band (e.g. $3 < z < 5$), their comoving 
density is $\simeq 1-2 \times 10^{-5} {\rm Mpc^{-3}}$, 
the same as the present-day number density of massive 
ellipticals of comparable mass to radio galaxies ($\ge 4 
L^{\star}$). 

For many years there has been growing evidence that the bulk of the
stellar populations in radio galaxies formed at high redshift, $z > 3$. I
conclude that the available data indicates that radio galaxies are
{\it not} special in this regard. Virtually all known 
luminous sub-mm sources are confined
to this same high-redshift regime, and their number density is sufficient 
to account for the formation of all massive ($\ge 4
L^{\star}$) ellipticals. These results suggest that CDM-based models need
to be tuned to produce very rapid collapse and conversion into stars of
the baryonic gas in the most massive haloes \cite{gra}. 

\section*{Acknowledgments}
I gratefully acknowledge the contributions of my collaborators on the 
projects covered in this review, especially Elese Archibald, Dave Hughes, 
Rob Ivison, Steve Rawlings, Omar Almaini, Dieter
Lutz, Chris Willott, Raul Jimenez, John Peacock, Suzie Scott and 
other members of the UK SCUBA Consortium.

\end{document}